\documentclass{Interspeech}

\interspeechcameraready

\title{
Source Verification for Speech Deepfakes
}

\author[affiliation={1}]{Viola}{Negroni}
\author[affiliation={1}]{Davide}{Salvi}
\author[affiliation={1}]{Paolo}{Bestagini}
\author[affiliation={1}]{Stefano}{Tubaro}

\affiliation{Dipartimento di Elettronica, Informazione e Bioingegneria}{Politecnico di Milano}{Italy}
\email{{viola.negroni, davide.salvi, paolo.bestagini, stefano.tubaro}@polimi.it}

\usepackage{comment}
\usepackage{tabularx}
\usepackage{longtable} 
\usepackage{colortbl}
\usepackage{multirow, multicol, array, makecell}
\usepackage{cleveref}
\usepackage{subcaption}

\crefname{section}{Section}{Sections}
\crefname{figure}{Figure}{Figures}
\crefname{table}{Table}{Tables}

\usepackage{glossaries}
\newacronym{cnn}{CNN}{Convolutional Neural Network}
\newacronym{dl}{DL}{Deep Learning}
\newacronym{ai}{AI}{Artificial Intelligence}
\newacronym{dnn}{DNN}{Deep Neural Network}
\newacronym{nn}{NN}{Neural Network}
\newacronym{mmf}{MMF}{MultiMedia Forensics}
\newacronym{vc}{VC}{Voice Conversion}
\newacronym{tts}{TTS}{Text-to-Speech}
\newacronym{ann}{ANN}{Artificial Neural Network}
\newacronym{df}{DF}{Deepfake}
\newacronym{ml}{ML}{Machine Learning}
\newacronym{mfcc}{MFCC}{Mel Frequency Cepstral Coefficient}
\newacronym{stft}{STFT}{Short Time Fourier Transform}
\newacronym{cqcc}{CQCC}{Constant Q Cepstral Coefficient}
\newacronym{roc}{ROC}{Receiver Operating Characteristic}
\newacronym{auc}{AUC}{Area Under the Curve}
\newacronym{tp}{TP}{True Positive}
\newacronym{tn}{TN}{True Negative}
\newacronym{fp}{FP}{False Positive}
\newacronym{fn}{FN}{False Negative}
\newacronym{tpr}{TPR}{True Positive Rate}
\newacronym{fpr}{FPR}{False Positive Rate}
\newacronym{tnr}{TNR}{True Negative Rate}
\newacronym{vae}{VAE}{Variational Autoencoders}
\newacronym{gan}{GAN}{Generative Adversarial Networks}
\newacronym{stlt}{STLT}{Short-Term and Long-Term}
\newacronym{gru}{GRU}{Gated recurrent unit}
\newacronym{ser}{SER}{Speech Emotion Recognition}
\newacronym{dft}{DFT}{Discrete Fourier Transform}
\newacronym{mae}{MAE}{Mean Absolute Error}
\newacronym{svm}{SVM}{Support Vector Machines}
\newacronym{gmm}{GMM}{Gaussian Mixture Models}
\newacronym{gnn}{GNN}{Graph Neural Network}
\newacronym{enf}{ENF}{Electric Network Frequency}
\newacronym{nlp}{NLP}{Natural Language Processing}
\newacronym{cm}{CM}{countermeasure}
\newacronym{mlp}{MLP}{Multi-Layer Perceptron}
\newacronym{ssl}{SSL}{Self-supervised Learning}
\newacronym{asv}{ASV}{Automatic Speaker Verification}
\newacronym{eer}{EER}{Equal Error Rate}
\newacronym{ir}{IR}{Impulse Response}
\newacronym{snr}{SNR}{Signal-to-Noise Ratio}

\begin{document}

\maketitle

\begin{abstract}

With the proliferation of speech deepfake generators, it becomes crucial not only to assess the authenticity of synthetic audio but also to trace its origin. While source attribution models attempt to address this challenge, they often struggle in open-set conditions against unseen generators. In this paper, we introduce the source verification task, which, inspired by speaker verification, determines whether a test track was produced using the same model as a set of reference signals. 
Our approach leverages embeddings from a classifier trained for source attribution, computing distance scores between tracks to assess whether they originate from the same source. We evaluate multiple models across diverse scenarios, analyzing the impact of speaker diversity, language mismatch, and post-processing operations. This work provides the first exploration of source verification, highlighting its potential and vulnerabilities, and offers insights for real-world forensic applications.

\end{abstract}

\section{Introduction}

The rapid advancement of AI-driven tools has made it easier than ever to create highly realistic synthetic content, including speech.
While these advances enable exciting applications in fields like entertainment, accessibility, and human-computer interaction, they also pose serious security and privacy risks by facilitating the diffusion of realistic falsified media for malicious purposes. 
This growing threat underscores the urgent need to develop robust methods for detecting and analyzing synthetic data, ensuring the integrity of digital communications.

Significant progress has been made in speech deepfake detection, with researchers leveraging deep learning and signal processing techniques to enhance classification accuracy~\cite{mo2022multi, ren2023lightweight, yang2024robust, cuccovillo2024audio, negroni2025leveraging, salvi2025freeze}. 
However, as detection models evolve, so do deepfake generation techniques, leading to an ongoing arms race between forensic researchers and adversarial actors.
Beyond merely distinguishing real from synthetic speech, a critical next step is to deepen our understanding of the content under analysis.
In this context, synthetic speech attribution which aims to identify the generative model behind a synthetic speech sample, has emerged as a valuable tool for forensic investigations and content authenticity verification.
Nevertheless, progress in this area has been limited, with current methods struggling with generalization, scalability, and performance in open-set scenarios.

Early approaches to speech deepfake attribution framed the problem as a closed-set classification task, where models were trained to identify the source of a synthetic sample from a predefined set of known generators~\cite{borrelli2021synthetic}. 
While effective in controlled settings, these methods struggle in open-set scenarios where deepfake samples may come from synthesizers that are not part of the training set~\cite{salvi2022exploring}.
Furthermore, closed-set models suffer from scalability issues: as the number of deepfake generators grows, updating these classifiers requires frequent retraining, a process that is computationally expensive and impractical given the rapid pace of innovation in the speech synthesis field.
These limitations make closed-set approaches unsuitable for real-world applications, where it is unrealistic to assume access to every possible generator.
Multiple challenges have been organized to address these issues, including ADD 2023~\cite{yi2023add} and SPCup 2022~\cite{salvi2023synthetic}, which introduced a dedicated track for speech deepfake attribution~\cite{zeng2023deepfake, lu2023detecting, tian2023deepfake}

An alternative approach to the attribution task focuses on analyzing the characteristics of the synthesis process rather than identifying the specific generator~\cite{zhu2022source, klein2024source}.
These methods extract features from various stages of the speech synthesis pipeline, such as acoustic models and vocoders, enabling better generalization across different deepfake generators by identifying common patterns and shared components within multiple synthesis architectures.
However, as end-to-end deepfake generators become increasingly sophisticated and integrate multiple synthesis steps, it is important to reconsider whether this approach remains a viable solution for source attribution.

In this paper, we address these limitations by introducing a novel framework for synthetic speech analysis: reframing deepfake attribution as a \textit{source verification} problem.
Inspired by \gls{asv}, where the goal is to verify whether an utterance belongs to a claimed speaker, our method evaluates whether a query track was created by the same generative model as a given set of reference signals.
This concept, which has already demonstrated success in speech deepfake detection~\cite{pianese2022deepfake, pianese2024training}, holds significant promise for open-set scenarios, offering a more flexible solution to the challenges of source tracing.
To implement this approach, we extract embeddings from the input tracks using a model trained for source attribution.
During inference, we compute similarity scores between test samples and reference tracks to determine whether they share the same generator.
We evaluate our method across diverse conditions, assessing its robustness against speaker diversity, language mismatches, and post-processing operations.

The paradigm shift we propose offers several advantages. 
First, it enables implicit attribution, removing the need for exhaustive training on all possible generators and improving generalization to unseen models.
Second, it enhances scalability, as new reference sets from emerging generators can be integrated without requiring system retraining, ensuring that the method remains effective as speech generation techniques evolve.
By reframing deepfake attribution as a verification task, we aim to overcome the limitations of existing approaches and provide a more robust solution for real-world forensic applications.
\section{Proposed Method}


\subsection{Problem Formulation}

The source verification task is formally defined as follows.  
Let $\mathfrak{R} = \{ \mathbf{x}_0, \mathbf{x}_1, \dots, \mathbf{x}_{R-1} \}$ represent a reference set of $R$ synthetic speech samples, all generated by the same speech synthesis model.
Given a test track $\mathbf{x}$ produced by an unknown generator, we aim to determine whether $\mathbf{x}$ originates from the same generator as the samples in $\mathfrak{R}$.
To this end, the test sample $\mathbf{x}$ is associated with a binary label $y \in \{0, 1\}$, where $y = 1$ indicates that $\mathbf{x}$ has been generated by the same source as the samples in $\mathfrak{R}$ and $y = 0$ indicates otherwise.  
The goal of the task is to predict the label $y$ for any given test sample $\mathbf{x}$.

\subsection{Proposed System}
\label{subsec:proposed_system}

\begin{figure}
    \centering
    \includegraphics[width=0.92\columnwidth]{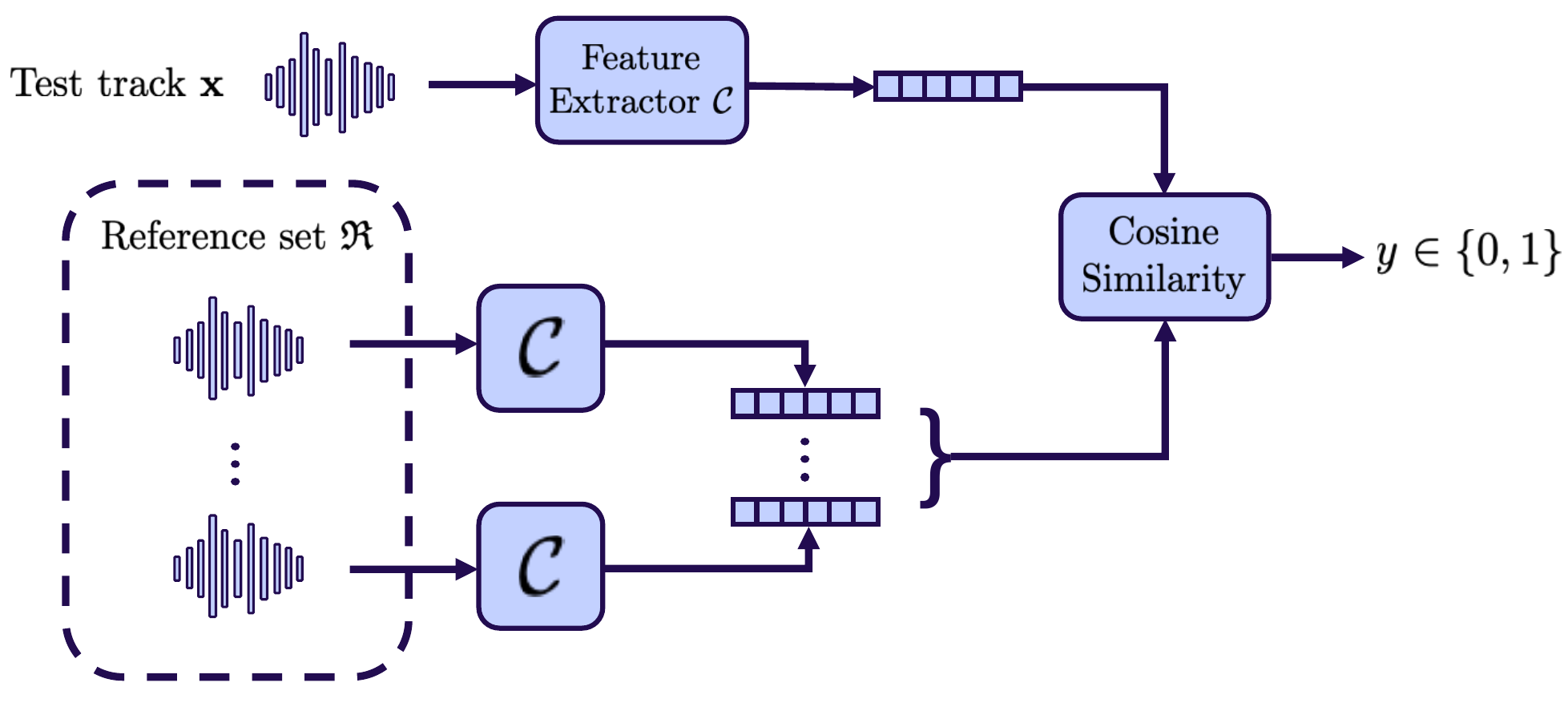}
    \caption{Pipeline of the proposed source verification method for speech deepfakes. A Feature Extractor $\mathcal{C}$ is leveraged as an embedding extractor for both the test track and the tracks in the reference set. Then, the cosine similarity is computed between the embedding of the test track and each reference embedding. The decision is made by taking the maximum similarity.}
    \label{fig:pipeline}
\end{figure}

In this paper, we introduce the novel task of source verification for speech deepfakes, designed to address the limitations of traditional synthetic speech attribution methods.
Our approach follows the multi-stage pipeline shown in \Cref{fig:pipeline}, which consists of the following key steps. \vspace{0.1em}

\noindent \textbf{Model Training.} We train a classifier $\mathcal{C}$ using a multi-class classification setup to perform synthetic speech attribution on a set of $N$ generators~\cite{salvi2022exploring}. This step ensures that the model learns to distinguish different synthetic speech sources, enabling it to produce meaningful feature vectors that capture information about the generative model used to create a given speech deepfake sample. \vspace{0.1em}

\noindent \textbf{Reference Set Construction.}
We construct a reference set $\mathfrak{R}$ using $R$ synthetic speech samples, all generated by the same speech synthesis model.
This generator is unseen during training, meaning that it is not included in the $N$ models on which $\mathcal{C}$ was trained.
This setup reflects a real-world scenario, where forensic systems must be able to operate in open-set conditions, i.e., generalize to deepfake models that were not encountered during training.
Using the classifier $\mathcal{C}$ as a feature extractor, we extract a feature vector $\mathbf{f}$ from each track in $\mathfrak{R}$, as in $\mathbf{f}_\mathbf{x} = \mathcal{C}(\mathbf{x})$. 
This is done by skipping the final decision layer of $\mathcal{C}$ and extracting embeddings from its last dense layer.
These feature vectors are then used to construct the reference feature set $\mathfrak{F}$, which has the same dimensionality of $\mathfrak{R}$.
Since $\mathcal{C}$ has been trained on the synthetic speech attribution task, we assume that the embeddings $\mathbf{f}$ will encode meaningful information about the model used to generate them. \vspace{0.1em}

\noindent \textbf{Verification Stage.} For a given test track $\mathbf{x}$, we extract its corresponding feature vector $\mathbf{f}_\mathbf{x}$ using $\mathcal{C}$ as a feature extractor.
Then, we compute the cosine similarity between $\mathbf{f}_\mathbf{x}$ and each feature vector in the reference set $\mathfrak{F}$ to determine whether $\mathbf{x}$ was generated by the same speech synthesis model used to produce the tracks in $\mathfrak{R}$. 
Following the approach motivated in~\cite{pianese2024training}, we use the maximum cosine similarity as the final decision statistic. \vspace{0.1em}

Our proposed source verification framework offers multiple advantages over traditional synthetic speech attribution models.
First, it naturally operates in open-set conditions, overcoming one of the primary limitations of conventional approaches that rely on a fixed set of known generators and struggle with unseen ones.
Our method, which relies only on the $N$ generators used to train $\mathcal{C}$, can generalize to any synthesis model, making it highly adaptable to real-world forensic scenarios and evolving deepfake technologies.
Second, rather than performing explicit attribution, our approach uses comparative verification, relying on distance measurements in the learned feature space. This enhances discrimination performance, improving robustness and accuracy in detecting speech deepfake sources.

\section{Experimental Setup}

In this section, we outline the experimental setup we used in our analyses. 
We first introduce the datasets used in our experiments, followed by a description of the models employed as classifiers $\mathcal{C}$ and their training strategies. 
Finally, we detail the methodology used for source verification.

\subsection{Datasets}
We evaluate the proposed framework on four different speech deepfake datasets to thoroughly assess its generalization capabilities.
All audio data are sampled at $f_\text{s} = \SI{16}{\kilo\hertz}$.\vspace{0.1em} 

\noindent\textbf{MLAAD \cite{muller2024mlaad} ($D_\text{MLA}$)}.
This is a large-scale dataset that includes synthetic speech signals generated using \num{82} \gls{tts} models across \num{33} distinct architectures, designed to evaluate anti-spoofing systems in multi-languages and multi-generators scenarios.
We use the fifth version of this corpus, which contains \num{378} hours of synthetic speech in \num{38} different languages, and follow the source tracing protocols proposed in~\cite{UsingMLAADforSourceTracing}. 
This dataset serves as the primary training corpus for our experiments. Unless otherwise specified, model training is conducted using a combination of the $D_\text{MLA}$ training and development partitions, which encompass \num{41} synthetic speech generators. Evaluation is performed on a separate test set, enabling open-set analysis.
\vspace{0.1em}

\noindent\textbf{ASVspoof 2019 \cite{todisco2019asvspoof} ($D_\text{ASV}$)}.
This dataset was developed for the homonymous challenge, aiming to push research towards the development of more effective \gls{asv} systems. We consider the LA-\textit{eval} partition of this corpus, which includes synthetic speech signals generated by 13 different synthetic speech generators. \vspace{0.1em}

\noindent\textbf{TIMIT-TTS \cite{salvi2023timit} ($D_\text{TIM}$)}.
This is a synthetic speech dataset derived from the VidTIMIT corpus~\cite{sanderson2002vidtimit}. For our experiments, we use its \textit{clean} partition, which contains speech signals generated using \num{12} different \gls{tts} models, with \num{7} dedicated to single-speaker synthesis and \num{5} supporting multiple speakers.
The single-speaker partition ($D_\text{TIM, SS}$) replicates the voice of Linda Johnson from LJSpeech~\cite{ljspeech17}, while the multi-speaker partition ($D_\text{TIM, MS}$) synthesizes voices from LibriSpeech~\cite{panayotov2015librispeech}. \vspace{0.1em}

\noindent\textbf{ADD 2023 \cite{yi2023add} ($D_\text{ADD}$)}.
This is a synthetic speech dataset released for the homonymous challenge, featuring tracks from Chinese speakers generated by \num{6} different speech deepfake generators.
In our experiments, we consider the Track 3-\textit{dev} partition of this corpus, which focuses on identifying the deepfake generator used to synthesize them.

\subsection{Considered Models and Training Strategy}

In our experimental analysis, we employ four different models as classifiers $\mathcal{C}$ to provide a comprehensive evaluation of the source verification task and ensure that our results are not biased toward a single architecture.
All the systems are state-of-the-art speech deepfake detectors, adapted to the synthetic speech attribution task following the methodology proposed in~\cite{salvi2022exploring}. 

The models we used are a modified version of ResNet18~\cite{he2016deep}, trained on Mel-spectrogram inputs with a dropout layer for regularization and a \num{256}-unit ReLU layer for feature extraction; a LCNN model~\cite{wu2018light}, which also processes Mel-spectrograms and it is known for its efficiency; RawNet2~\cite{tak2021end}, an end-to-end model designed for raw audio processing that incorporates Sinc-filters~\cite{ravanelli2018speaker}; and AASIST~\cite{jung2022aasist}, a \gls{gnn} model integrating a RawNet2-based encoder, graph modules and an attention layer to model temporal and spectral artifacts in synthetic speech.

All models were trained using the same methodology. 
We trained them for the synthetic speech attribution task, aiming to detect which generator has been used to synthesize a given speech deepfake track.
We used 4-second input audio segments and adjusted the final dense layer of each model to match the number of classes in the classification task.
We trained the models until convergence by monitoring the value of the validation loss, computed via Cross-Entropy function.
We employed cosine annealing for learning rate scheduling and early stopping to prevent overfitting.
We assumed \num{10} epochs as early stopping, a batch size of \num{128}, and a learning rate $lr=$\num{e-4}.

\subsection{Source Verification Process}

During the verification stage, we leverage the classifiers $\mathcal{C}$ as feature extractors, obtaining representations from the last dense layer before the classification head.
These extracted features are informative for the speech attribution task, and serve as the basis to perform the comparison between test tracks and reference sets.
For all experiments, we use \num{5} reference tracks per test generator, a choice based on a similar ad hoc experiment as in \cite{pianese2024training}, omitted here for brevity.
During inference, the similarity between a test track $\mathbf{x}$ and the reference set $\mathfrak{R}$ of a given generator is determined by computing the maximum cosine similarity between the feature vector of $\mathbf{x}$ and those of all reference tracks, following the approach validated in~\cite{pianese2024training}.

As the entire pipeline relies on feature vector comparisons, the verification process occurs in the latent space learned by each classifier $\mathcal{C}$. 
Consequently, we expect verification performance to be strongly influenced by the dataset used for training, similar to what has been observed in the \gls{asv} domain~\cite{xia2019cross}. 
This aspect will be explored in detail in the following section.

\section{Results}
\label{sec:results} 

In this section, we evaluate the proposed pipeline on the source verification task. 
We assess its performance under various experimental conditions, analyzing its robustness to factors such as speaker diversity, language mismatches, and post-processing operations.

\subsection{Source Verification}
\label{subsec:results_task}

\begin{table}
\caption{EER and AUC values (\%) of the considered systems on the source verification task across multiple datasets.}
\label{tab:verification_results}
\centering
\resizebox{\columnwidth}{!}{
\begin{tabular}{ccccccc|cc}
\hline
\toprule
\multirow{2}{*}{} & \multicolumn{2}{c}{$D_\text{MLA}$} & \multicolumn{2}{c}{$D_\text{ASV}$} & \multicolumn{2}{c}{$D_\text{TIM}$} & \multicolumn{2}{c}{Average} \\ \cmidrule(lr){2-3} \cmidrule(lr){4-5} \cmidrule(lr){6-7} \cmidrule(lr){8-9}
                  & EER $\downarrow$ & AUC $\uparrow$  & EER $\downarrow$ & AUC $\uparrow$  & EER $\downarrow$ & AUC $\uparrow$ & EER $\downarrow$ & AUC $\uparrow$  \\ \midrule
ResNet18 & \textbf{4.8}     & \textbf{99.1}   & \textbf{25.5}    & \textbf{82.0}   &\textbf{8.7}      & \textbf{95.6}  & \textbf{13.0}  & \textbf{92.2}  \\
LCNN     & 15.0             & 92.4            & 27.0             & 80.9            & 16.7             & 89.6           & 19.6           & 87.6            \\
AASIST   & 8.3              & 97.4            & 27.6             & 80.4            & 11.8             & 93.9           & 15.9           & 90.6             \\
RawNet2  & 9.4              & 96.8            & 34.6             & 70.4            & 12.0             & 93.8           & 18.7           & 87.0              \\ \bottomrule   
\end{tabular}}
\end{table}

As an initial investigation, we assess the feasibility of the proposed source verification task, analyzing whether it is possible to verify the generator of a deepfake speech signal.
We do so by using four classifiers $\mathcal{C}$ as feature extractors, trained on the $D_\text{MLA}$ training set~\cite{UsingMLAADforSourceTracing} and evaluated in open-set conditions, meaning that all speech generators used in test are not included in the training set.
The test datasets that we consider are the $D_\text{MLA}$ test set partition~\cite{UsingMLAADforSourceTracing}, $D_\text{ASV}$ and $D_\text{TIM}$.

\Cref{tab:verification_results} presents the results of this analysis in terms of \gls{eer} and \gls{auc}.  
The best performance are obtained on the $D_\text{MLA}$ corpus, where \gls{auc} values remain consistently above \num{90}\%.
This strong performance is likely due to similarities in data processing pipelines between the training and test sets, as they both come from the same dataset, even though the speech generators are different.
For the other sets, performance on $D_\text{TIM}$ is comparable to that on $D_\text{MLA}$, while a noticeable performance drop is observed across all models on $D_\text{ASV}$. This may be attributed to differences in generator types: while both $D_\text{MLA}$ and $D_\text{TIM}$ contain only \gls{tts} generators, $D_\text{ASV}$ also includes \gls{vc} and hybrid generators, which could introduce additional complexity.
Overall, the best-performing model is ResNet18, achieving an average \gls{eer} of \num{13.0}\% and an \gls{auc} of \num{92.2}\%, indicating its effectiveness for the source verification task.
Therefore, we will consider only this model for the following experiments.

\subsection{Speakers Diversity Analysis}
\label{subsec:speakers_exp}

\begin{table}
\caption{EER and AUC values (\%) for ResNet18 models trained on single-speaker (R18-SS) and multi-speaker (R18-MS) data, evaluated on multi-speaker ($D_\text{MLA, Spk}$, $D_\text{ASV}$, and $D_\text{TIM, MS}$) and single-speaker ($D_\text{TIM, SS}$) data.}
\label{tab:speaker_results}
\resizebox{\columnwidth}{!}{
\centering
\begin{tabular}{ccccccccc}
\hline
\toprule
\multirow{2}{*}{}    & \multicolumn{2}{c}{\text{$D_\text{MLA, Spk}$}} & \multicolumn{2}{c}{\text{$D_\text{ASV}$}} & \multicolumn{2}{c}{\text{$D_\text{TIM, SS}$}} & \multicolumn{2}{c}{\text{$D_\text{TIM, MS}$}} \\ \cmidrule(lr){2-3} \cmidrule(lr){4-5} \cmidrule(lr){6-7} \cmidrule(lr){8-9}
                     & EER $\downarrow$    & AUC $\uparrow$     & EER $\downarrow$     & AUC $\uparrow$     & EER $\downarrow$     & AUC $\uparrow$        & EER $\downarrow$     & AUC $\uparrow$        \\ \midrule
R18-SS & 28.2                & 78.5               & 37.2                 & 67.9               & \textbf{28.0}        & \textbf{83.1}         & 22.4                 & 87.9                  \\
R18-MS & \textbf{18.6}       & \textbf{88.9}      & \textbf{28.1}        & \textbf{79.7}      & 31.6                 & 75.1                  & \textbf{12.7}        & \textbf{94.7}         \\   \bottomrule  
\end{tabular}
}
\vspace{-.5em}
\end{table}

In this experiment, we investigate how the source verification performance is affected by the distribution of speakers in the training data of the classifier $\mathcal{C}$.
This is a crucial aspect, as the presence of non-optimal speaker distribution in the training data may introduce bias, potentially leading the model to differentiate tracks based on speaker identity rather than the distinguishing characteristics of the speech generator~\cite{salvi2023timit}.
To explore this effect, we train two versions of ResNet18: \textit{R18-SS}, which is trained on \num{10} generators from $D_\text{MLA}$ that produce speech with a single voice (LJSpeech), and \textit{R18-MS}, that is trained on \num{10} multi-speaker $D_\text{MLA}$ generators.
Both models are evaluated on a set of \num{16} multi-speaker generators that were not seen during training ($D_\text{MLA, Spk}$), as well as on the $D_\text{ASV}$, $D_\text{TIM, SS}$, and $D_\text{TIM, MS}$ datasets.
To ensure consistency, we restrict all $D_\text{MLA, Spk}$ data to English-language samples.

\Cref{tab:speaker_results} presents the results of this analysis.
R18-MS consistently outperforms R18-SS across all multi-speaker datasets. However, its performance drops on $D_\text{TIM, SS}$, the only single-speaker test dataset that we considered.
These findings underscore the critical role of speaker-specific features in synthetic speech attribution and, consequently, in source verification.
Models trained on multi-speaker data, such as R18-MS, may inadvertently rely on speaker-related cues rather than generator-specific characteristics. In contrast, single-speaker models, such as R18-SS, which are trained to focus solely on generator-specific features, demonstrate greater robustness in more challenging scenarios, such as $D_\text{TIM, SS}$, where speaker identity is not a distinguishing factor.


\subsection{Languages Analysis}
\label{subsec:languages_exp}

\begin{table}
\centering
\caption{EER and AUC values (\%) for ResNet18 models trained on English-only (R18-EL) and non-English (R18-ML) data, evaluated on English ($D_\text{MLA, EL}$, $D_\text{ASV}$, $D_\text{TIM}$) and non-English ($D_\text{MLA, ML}$, $D_\text{ADD}$) data.}
\label{tab:languages_results}
\resizebox{.6\columnwidth}{!}{
\begin{tabular}{@{}lcccc@{}}
\toprule
                       & \multicolumn{2}{c}{R18-EL} & \multicolumn{2}{c}{R18-ML} \\ \cmidrule(lr){2-3} \cmidrule(lr){4-5}
                       & EER $\downarrow$  & AUC $\uparrow$            & EER $\downarrow$  & AUC $\uparrow$               \\ \midrule
$D_\text{MLA, EL}$       & \textbf{18.2}           & \textbf{90.4}          & 21.0             & 86.7             \\
$D_\text{MLA, ML}$       & 4.5            & 99.1          & \textbf{3.0}              & \textbf{99.4}             \\
$D_\text{ASV}$           & \textbf{26.6}           & \textbf{81.7}          & 28.9             & 77.9             \\
$D_\text{TIM}$           & 12.5           & 92.8          & \textbf{12.0}             & \textbf{93.0}              \\
$D_\text{ADD}$           & 19.7           & 89.3          & \textbf{13.4}             & \textbf{93.7}             \\ \bottomrule
\end{tabular}}
\vspace{-0.5em}
\end{table}

In this experiment, we investigate whether the language of the training data influences the discrimination capabilities of the classifier.
Specifically, we aim to determine whether training on a single language or multiple languages leads to better performance in different scenarios for the source verification task.

We do so by training two ResNet18 models: \textit{R18-EL}, trained on \num{13} English-only generators, and \textit{R18-ML}, trained on \num{13} generators spanning various non-English languages.
For evaluation, both models are tested on two distinct $D_\text{MLA}$ subsets: $D_\text{MLA, EL}$, that contains signals from \num{9} English-only generators, and $D_\text{MLA, ML}$, which includes data from \num{9} non-English generators, each in a language not present in the training data.
To ensure a fair comparison, we manually curated the training and test subsets, minimizing generator overlap and maintaining balance across all partitions.
Additionally, based on the findings of the previous experiment, all $D_\text{MLA}$ data used in this study come from multi-speaker generators.
We also evaluate both models on $D_\text{ASV}$ and $D_\text{TIM}$ to assess their ability to generalize across additional English-speaking datasets, as well as on $D_\text{ADD}$ for non-English (Chinese) data.

\Cref{tab:languages_results} presents the results of this analysis.
Our findings indicate that R18-EL performs better on English datasets ($D_\text{MLA, EL}$ and $D_\text{ASV}$), while R18-ML outperforms R18-EL on non-English datasets ($D_\text{MLA, ML}$ and $D_\text{ADD}$).
Performances on $D_\text{TIM}$ are comparable between the two models.
It is worth noting that the two systems models have excellent source verification performance on $D_\text{MLA, ML}$ than on $D_\text{MLA, EL}$. 
This can likely be attributed to some degree of generator overlap between training and test sets, despite the language differences.
This suggests that certain generator-specific features are transferable across languages, reinforcing the idea that source verification models may benefit from a broader linguistic diversity in training.

\subsection{Robustness Analysis}
\label{subsec:robustness_exp}

As a final experiment, we evaluate the robustness of the proposed framework against a wide range of post-processing operations.
This is a crucial analysis because the predictions of the classifiers $\mathcal{C}$ rely on subtle, nearly inaudible fingerprints embedded by speech deepfake generators.
Post-processing operations that hinder these traces could significantly degrade source verification performance, and we aim to quantify this potential drop. 
The post-processing operations we evaluate include Gaussian noise injection applied with a randomly selected \gls{snr} between \num{25} and \SI{15}{\decibel}; MP3 compression at randomly chosen bitrates (16, 32, 64 and 128 kbps); speech enhancement performed using the SepFormer model \cite{subakan2021attention}; convolution with \glspl{ir} from various mobile phones to simulate different recording conditions~\cite{salvi2025poliphone}.
For this evaluation, we use the same ResNet18 model used in \Cref{subsec:results_task}, and test it on $D_\text{MLA}$ test, $D_\text{ASV}$ and $D_\text{TIM}$.

\Cref{tab:robustness_results} presents the results of this analysis.
Among all post-processing operations, speech enhancement has the most significant impact on performance, causing an average \gls{eer} increase of over \num{25}\%.
We hypothesize that this is because the ResNet18 model relies on generator-specific artifacts that are unrelated to speech content.
By enhancing the intelligibility of speech, the enhancement model may simultaneously suppress these forensic traces, making the source verification task more challenging.
Conversely, convolution with mobile phone \glspl{ir} has a less significant effect, reducing the average \gls{auc} by less than \num{10}\%.
This suggests that reverberation and acoustic transformations do not completely mask the generator-specific characteristics, allowing the signals to preserve much of their discriminative information for source attribution.
Finally, noise injection and MP3 compression exhibit a moderate and comparable impact on performance, indicating that while these operations introduce distortions, they do not fully hinder the generator fingerprints.
These findings highlight the vulnerability of source verification models to certain types of post-processing operations, underscoring the importance of considering real-world conditions when designing and evaluating source verification frameworks.

\begin{table}
\caption{EER and AUC values (\%) for ResNet18 on clean and post-processed data. Best values are in bold, worst in italics.}
\label{tab:robustness_results}
\resizebox{\columnwidth}{!}{
\centering
\begin{tabular}{ccccccc|cc}
\toprule
\multirow{2}{*}{} & \multicolumn{2}{c}{\text{$D_\text{MLA}$}} & \multicolumn{2}{c}{\text{$D_\text{ASV}$}} & \multicolumn{2}{c}{\text{$D_\text{TIM}$}} & \multicolumn{2}{c}{Average}\\  \cmidrule(lr){2-3} \cmidrule(lr){4-5} \cmidrule(lr){6-7} \cmidrule(lr){8-9} 
                    & EER $\downarrow$ & AUC $\uparrow$ & EER $\downarrow$ & AUC $\uparrow$ & EER $\downarrow$ & AUC $\uparrow$ & EER $\downarrow$ & AUC $\uparrow$\\ \midrule
Clean                                    & 4.8           & 99.1          & 25.5          & 82.0          & 8.7           & 95.6          & 13.0          & 92.2 \\ \midrule
Noise Inj.                               & 19.0          & 89.4          & 36.7          & 68.2          & 19.8          & 88.8          & 25.5          & 82.1 \\
MP3 Compr.                               & \textbf{15.0} & \textbf{92.3} & \textit{39.7} & \textit{64.2} & \textbf{15.4} & \textbf{92.1} & \textbf{23.4} & \textbf{82.9} \\ 
Speech Enh.                              & \textit{37.6} & \textit{67.9} & \textbf{31.8} & \textbf{73.9} & \textit{52.4} & \textit{46.6} & \textit{40.6} & \textit{62.8} \\
IR Conv.                                 & 19.2          & 88.9          & 34.1          & 71.8          & 18.0          & 89.3          & 23.8          & 83.3 \\ \bottomrule
\end{tabular}}
\vspace{-0.5em}
\end{table}
\section{Conclusions}

In this paper, we introduced the source verification task, a novel approach to source tracing for speech deepfakes.
Inspired by \gls{asv}, the proposed method determines whether a query track was generated by the same model as a set of reference samples, eliminating the need for exhaustive training on every possible generator.
Our evaluation demonstrates strong generalization of source verification to unseen models, while addressing scalability challenges and examining limitations related to speaker diversity, language mismatches, and the impact of post-processing operations.
Unlike traditional attribution methods, this approach enables implicit attribution, making it more adaptable to evolving speech deepfake technologies.
This work lays the foundation for more scalable and practical forensic tools for source tracing.
Future research could explore alternative approaches to embedding extraction, possibly 
leveraging decomposition techniques for a more robust and interpretable subspace modeling \cite{kuhn1998eigenvoices}, further analyzing how performance may be affected by the choice of training data classes, and experimenting with more sophisticated decision rules.

\section{Acknowledgements}
This work was supported by the FOSTERER project, funded by the Italian Ministry of Education, University, and Research within the PRIN 2022 program.
This work was partially supported by the European Union - Next Generation EU under the Italian National Recovery and Resilience Plan (NRRP), Mission 4, Component 2, Investment 1.3, CUP D43C22003080001, partnership on “Telecommunications of the Future” (PE00000001 - program “RESTART”).
This work was partially supported by the European Union - Next Generation EU under the Italian National Recovery and Resilience Plan (NRRP), Mission 4, Component 2, Investment 1.3, CUP D43C22003050001, partnership on ``SEcurity and RIghts in the CyberSpace’’ (PE00000014 - program ``FF4ALL-SERICS’’).

\bibliographystyle{IEEEtran}
\bibliography{bibliography}

\begin{thebibliography}{10}
\providecommand{\url}[1]{#1}
\csname url@samestyle\endcsname
\providecommand{\newblock}{\relax}
\providecommand{\bibinfo}[2]{#2}
\providecommand{\BIBentrySTDinterwordspacing}{\spaceskip=0pt\relax}
\providecommand{\BIBentryALTinterwordstretchfactor}{4}
\providecommand{\BIBentryALTinterwordspacing}{\spaceskip=\fontdimen2\font plus
\BIBentryALTinterwordstretchfactor\fontdimen3\font minus \fontdimen4\font\relax}
\providecommand{\BIBforeignlanguage}[2]{{%
\expandafter\ifx\csname l@#1\endcsname\relax
\typeout{** WARNING: IEEEtran.bst: No hyphenation pattern has been}%
\typeout{** loaded for the language `#1'. Using the pattern for}%
\typeout{** the default language instead.}%
\else
\language=\csname l@#1\endcsname
\fi
#2}}
\providecommand{\BIBdecl}{\relax}
\BIBdecl

\bibitem{mo2022multi}
Y.~Mo and S.~Wang, ``Multi-task learning improves synthetic speech detection,'' in \emph{IEEE International Conference on Acoustics, Speech and Signal Processing (ICASSP)}, 2022.

\bibitem{ren2023lightweight}
Y.~Ren, H.~Peng, L.~Li, and Y.~Yang, ``Lightweight voice spoofing detection using improved one-class learning and knowledge distillation,'' \emph{IEEE Transactions on Multimedia}, 2023.

\bibitem{yang2024robust}
Y.~Yang, H.~Qin, H.~Zhou, C.~Wang, T.~Guo, K.~Han, and Y.~Wang, ``A robust audio deepfake detection system via multi-view feature,'' in \emph{IEEE International Conference on Acoustics, Speech and Signal Processing (ICASSP)}, 2024.

\bibitem{cuccovillo2024audio}
L.~Cuccovillo, M.~Gerhardt, and P.~Aichroth, ``Audio transformer for synthetic speech detection via formant magnitude and phase analysis,'' in \emph{IEEE International Conference on Acoustics, Speech and Signal Processing (ICASSP)}, 2024.

\bibitem{negroni2025leveraging}
V.~Negroni, D.~Salvi, A.~I. Mezza, P.~Bestagini, and S.~Tubaro, ``Leveraging mixture of experts for improved speech deepfake detection,'' in \emph{IEEE International Conference on Acoustics, Speech and Signal Processing (ICASSP)}, 2025.

\bibitem{salvi2025freeze}
D.~Salvi, V.~Negroni, L.~Bondi, P.~Bestagini, and S.~Tubaro, ``Freeze and learn: Continual learning with selective freezing for speech deepfake detection,'' in \emph{IEEE International Conference on Acoustics, Speech and Signal Processing (ICASSP)}, 2025.

\bibitem{borrelli2021synthetic}
C.~Borrelli, P.~Bestagini, F.~Antonacci, A.~Sarti, and S.~Tubaro, ``Synthetic speech detection through short-term and long-term prediction traces,'' \emph{EURASIP Journal on Information Security}, vol. 2021, 2021.

\bibitem{salvi2022exploring}
D.~Salvi, P.~Bestagini, and S.~Tubaro, ``Exploring the synthetic speech attribution problem through data-driven detectors,'' in \emph{IEEE International Workshop on Information Forensics and Security (WIFS)}, 2022.

\bibitem{yi2023add}
J.~Yi, J.~Tao, R.~Fu, X.~Yan, C.~Wang, T.~Wang, C.~Y. Zhang, X.~Zhang, Y.~Zhao, Y.~Ren \emph{et~al.}, ``Add 2023: the second audio deepfake detection challenge,'' in \emph{DADA@ IJCAI}, 2023.

\bibitem{salvi2023synthetic}
D.~Salvi, C.~Borrelli, P.~Bestagini, F.~Antonacci, M.~Stamm, L.~Marcenaro, and A.~Majumdar, ``Synthetic speech attribution: Highlights from the ieee signal processing cup 2022 student competition [sp competitions],'' \emph{IEEE Signal Processing Magazine}, vol.~40, no.~6, pp. 92--98, 2023.

\bibitem{zeng2023deepfake}
X.-M. Zeng, J.-T. Zhang, K.~Li, Z.-L. Liu, W.-L. Xie, and Y.~Song, ``Deepfake algorithm recognition system with augmented data for add 2023 challenge.'' in \emph{DADA@ IJCAI}, 2023.

\bibitem{lu2023detecting}
J.~Lu, Y.~Zhang, Z.~Li, Z.~Shang, W.~Wang, and P.~Zhang, ``Detecting unknown speech spoofing algorithms with nearest neighbors.'' in \emph{DADA@ IJCAI}, 2023.

\bibitem{tian2023deepfake}
Y.~Tian, Y.~Chen, Y.~Tang, and B.~Fu, ``Deepfake algorithm recognition through multi-model fusion based on manifold measure.'' in \emph{DADA@ IJCAI}, 2023.

\bibitem{zhu2022source}
T.~Zhu, X.~Wang, X.~Qin, and M.~Li, ``Source tracing: detecting voice spoofing,'' in \emph{IEEE Asia-Pacific Signal and Information Processing Association Annual Summit and Conference (APSIPA ASC)}, 2022.

\bibitem{klein2024source}
N.~Klein, T.~Chen, H.~Tak, R.~Casal, and E.~Khoury, ``Source tracing of audio deepfake systems,'' in \emph{Proc. INTERSPEECH}, 2024.

\bibitem{pianese2022deepfake}
A.~Pianese, D.~Cozzolino, G.~Poggi, and L.~Verdoliva, ``Deepfake audio detection by speaker verification,'' in \emph{IEEE International Workshop on Information Forensics and Security (WIFS)}, 2022.

\bibitem{pianese2024training}
------, ``Training-free deepfake voice recognition by leveraging large-scale pre-trained models,'' in \emph{Proc. ACM Workshop on Information Hiding and Multimedia Security}, 2024.

\bibitem{muller2024mlaad}
N.~M. M{\"u}ller, P.~Kawa, W.~H. Choong, E.~Casanova, E.~G{\"o}lge, T.~M{\"u}ller, P.~Syga, P.~Sperl, and K.~B{\"o}ttinger, ``{MLAAD: The Multi-Language Audio Anti-Spoofing Dataset},'' \emph{IEEE International Joint Conference on Neural Networks (IJCNN)}, 2024.

\bibitem{UsingMLAADforSourceTracing}
N.~M{\"u}ller, ``Using mlaad for source tracing of audio deepfakes,'' \url{https://deepfake-total.com/sourcetracing}, Fraunhofer AISEC, 11 2024.

\bibitem{todisco2019asvspoof}
M.~Todisco, X.~Wang, V.~Vestman, M.~Sahidullah, H.~Delgado, A.~Nautsch, J.~Yamagishi, N.~Evans, T.~Kinnunen, and K.~A. Lee, ``{ASVspoof 2019: Future horizons in spoofed and fake audio detection},'' in \emph{{Proc. INTERSPEECH}}, 2019.

\bibitem{salvi2023timit}
D.~Salvi, B.~Hosler, P.~Bestagini, M.~C. Stamm, and S.~Tubaro, ``{TIMIT-TTS: a Text-to-Speech Dataset for Multimodal Synthetic Media Detection},'' \emph{IEEE Access}, 2023.

\bibitem{sanderson2002vidtimit}
C.~Sanderson, ``{The VidTIMIT database},'' IDIAP, Tech. Rep., 2002.

\bibitem{ljspeech17}
K.~Ito and L.~Johnson, ``{The LJ Speech Dataset},'' \url{https://keithito.com/LJ-Speech-Dataset/}, 2017.

\bibitem{panayotov2015librispeech}
V.~Panayotov, G.~Chen, D.~Povey, and S.~Khudanpur, ``Librispeech: an asr corpus based on public domain audio books,'' in \emph{IEEE International Conference on Acoustics, Speech and Signal Processing (ICASSP)}, 2015.

\bibitem{he2016deep}
K.~He, X.~Zhang, S.~Ren, and J.~Sun, ``Deep residual learning for image recognition,'' in \emph{IEEE conference on Computer Vision and Pattern Recognition (CVPR)}, 2016.

\bibitem{wu2018light}
X.~Wu, R.~He, Z.~Sun, and T.~Tan, ``{A light CNN for deep face representation with noisy labels},'' \emph{{IEEE Transactions on Information Forensics and Security}}, vol.~13, no.~11, 2018.

\bibitem{tak2021end}
H.~Tak, J.~Patino, M.~Todisco, A.~Nautsch, N.~Evans, and A.~Larcher, ``End-to-end anti-spoofing with rawnet2,'' in \emph{IEEE International Conference on Acoustics, Speech and Signal Processing (ICASSP)}, 2021.

\bibitem{ravanelli2018speaker}
M.~Ravanelli and Y.~Bengio, ``Speaker recognition from raw waveform with sincnet,'' in \emph{IEEE spoken language technology workshop (SLT)}, 2018.

\bibitem{jung2022aasist}
J.-w. Jung, H.-S. Heo, H.~Tak, H.-j. Shim, J.~S. Chung, B.-J. Lee, H.-J. Yu, and N.~Evans, ``Aasist: Audio anti-spoofing using integrated spectro-temporal graph attention networks,'' in \emph{IEEE international conference on acoustics, speech and signal processing (ICASSP)}, 2022.

\bibitem{xia2019cross}
W.~Xia, J.~Huang, and J.~H. Hansen, ``Cross-lingual text-independent speaker verification using unsupervised adversarial discriminative domain adaptation,'' in \emph{IEEE International Conference on Acoustics, Speech and Signal Processing (ICASSP)}, 2019.

\bibitem{subakan2021attention}
C.~Subakan, M.~Ravanelli, S.~Cornell, M.~Bronzi, and J.~Zhong, ``Attention is all you need in speech separation,'' in \emph{IEEE International Conference on Acoustics, Speech and Signal Processing (ICASSP)}, 2021.

\bibitem{salvi2025poliphone}
D.~Salvi, D.~U. Leonzio, A.~Giganti, C.~Eutizi, S.~Mandelli, P.~Bestagini, and S.~Tubaro, ``Poliphone: A dataset for smartphone model identification from audio recordings,'' \emph{IEEE Access}, 2025.

\bibitem{kuhn1998eigenvoices}
R.~Kuhn, P.~Nguyen, J.-C. Junqua, L.~Goldwasser, N.~Niedzielski, S.~Fincke, K.~L. Field, and M.~Contolini, ``Eigenvoices for speaker adaptation.'' in \emph{ICSLP}, vol.~98, 1998, pp. 1774--1777.

\end{thebibliography}

\end{document}